\documentstyle[twocolumn,epsfig,psfig]{article}
\def\vt{\vartheta}
\def\be{\begin{equation}}
\def\ee{\end{equation}}

\begin{document}

\title{Universal conductance fluctuations in non-integer dimensions}

\author{I.Trav\v enec$^*$\\
Institute of Physics, Slovak Academy of Sciences, D\'ubravsk\'a cesta 9,
842 28 Bratislava, Slovakia}

\maketitle

\begin{abstract} 

We propose an Ansatz for universal conductance fluctuations in continuous
dimensions from 0 up to 4. The Ansatz agrees with known formulas for integer 
dimensions 1, 2 and 3, both for hard wall and periodic boundary conditions. 
The method is based solely on the knowledge of energy spectrum and standard
assumptions.
We also study numerically the conductance fluctuations in 4D Anderson model, 
depending on system size $L$ and disorder $W$. We find a small plateau with 
a value diverging logarithmically with increasing $L$. 
Universality gets lost just in 4D.

\end{abstract}

\bigskip


\bigskip
Disordered systems usually possess the metallic regime, where Ohm's law
for mean conductance $\langle g \rangle$ works well at least for
cubic samples and the distribution 
of $g$ for various realizations of the same disorder is Gaussian with
constant (disorder, mean free path $l_e$ and $\langle g \rangle$ independent) 
width, called universal conductance fluctuations (UCF). Higher cummulants 
of $g$ should disappear as some power of $1/\langle g \rangle$ \cite{AKL1}, 
though recent experiment on gold wire did not confirm this in quasi-1D 
\cite{Moh}. Recently \cite{nas} we analysed the statistical properties 
of conductance on bifractal lattices \cite{Gruss}. It became clear, that  
$\langle g \rangle$ and $var\ g$ depend (besides the spectral dimension 
$d_s$) on lattice topology.
Simply speaking, bifractals are no hypercubes. But we hope that by changing 
other parameters, say anisotropy, we can tune the systems to cubic-like. 
If the UCF of anisotropic bifractals become those of non-integer dimensional 
hypercubes, other parameters of these systems may be comparable.
The main goal of this work is to find a way to calculate the UCF also
for non-integer dimensions. Rewriting (\ref{varg}) as a simple integral,
we will propose its analytical continuation to real dimensions.

Two (on first sight) different expressions were given for UCF in literature 
\cite{LSF} and \cite{vanR}. Let us comment this seeming ambiguity. 
In classical papers of Lee, Stone and Fukuyama \cite{LSF} - Appendix, formulas 
forUCF in 3D (2D, 1D) were given as a sum of three diagrammatic terms $F_a, 
F_b$ and $ F_c$. They can be written in terms of convolutions, e. g. (the simple 
one-loop "Meeron" diagram):
\be\label{bubble}
F_a = {2\over L^4} \int \Pi({\bf r},{\bf r'}) \Pi({\bf r'},{\bf r}) 
d{\bf r} d{\bf r}'
\ee
where $\Pi({\bf r},{\bf r'})$ is standard propagator in a box, $F_b$ and 
$F_c$ contain 3 and 4 propagators (cyclically in one loop), respectively.
In 1D it is possible to show analytically, that $F_b= -F_a$ 
and $F_c=3/4\ F_a$, with some tricky cancelling of divergences, 
Ref. \cite{vRN}. Numerical calculations show, that the same holds for 2D and 3D. 
Thus we arrive at the formula, given in Ref. \cite{vanR} and representing 
$3/4\ F_a$ from Ref. \cite{LSF}:
\be\label{varg}
\langle g^2 \rangle _c = {12 \over \pi^4} \sum_{i_z=1}^{\infty} 
\sum_{i_x=0}^{\infty} \sum_{i_y=0}^{\infty} (i_z^2+i_x^2+i_y^2)^{-2}
\ee
where the number of sums defines the dimensionality, hard wall boundary 
conditions are applied and we limit ourselves to cubes ($L_x=L_y=L_z$). 
This relation was derived using the technique of Hikami boxes. It gives 
in one dimension
\be\label{U1D}
\langle g^2 \rangle _c^{1D} ={12\over\pi^4 }\ \zeta (4)
= {2 \over 15}
\ee
$\zeta(x)$ is Rieman's function. No such simple formulas were
given for higher dimensions yet. It is also clear, that a four-fold sum,
i. e. 4D case of Eq. (\ref{varg}) diverges and $d=4$ plays a role of some kind "critical 
dimension". According to the remark 6 of Ref. \cite{AKL}, it is possible
that also ergodicity gets lost at that point.

\medskip

Dealing with diffusive part of conductivity itself, the following formula 
for diagonal part of the propagator in 3D was given \cite{Braun} and \cite{Nik}:
\be\label{pirr}
\Pi({\bf r},{\bf r}) = -{\delta g \over g} = {2\over g\pi^2} S(y)
\ee
where $\delta g$ is the diffusive part of conductivity and a function
$S(y)$ was specified as follows:
\be\label{HW}
S_{\rm HW}(y)=\sum_{i_z=1}^{\infty}\sum_{i_x,i_y=0}^{\infty}
{\exp [-\pi (i_x^2+i_y^2+i_z^2)y]\over i_x^2+i_y^2+i_z^2}
\ee
\be\label{PBC}
S_{\rm PBC}(y)=\sum_{i_z=1}^{\infty}\sum_{i_x,i_y=-\infty}^{\infty}
{\exp [-\pi (4 i_x^2+4 i_y^2+i_z^2)y]\over 4 i_x^2+4 i_y^2+i_z^2}.
\ee
Here $y=\pi\phi_d(l_e/L)^2$, $\phi_d$ is dimension 
specific, e. g. 1/3 in 3D, but for our purposes insignificant constant. 
HW stands for hard wall and PBC for periodic boundary conditions in directions
perpendicular to the current, flowing in the $z$ direction. Later we will
give an alternative meaning to the variable $y$.
Contrary to Ref. \cite{Braun} we include zero modes in our Eq. (\ref{PBC}), 
$i_x=0$ and $i_y=0$, similarly to Ref. \cite{Mir}. Our normalization of $y$ 
differs slightly from that in Ref. \cite{Braun}. 
Instead of complete elliptic integrals, we will make use 
of $\vt_3$ from the family of Jacobi elliptic functions, defined as 
\cite{GR}:
\be\label{deft}
\vt_3(u,q)=1+2\sum_{n=1}^{\infty}q^{n^2}\cos(2nu)
\ee
where we set $q=\exp(-\pi y)$ and $u=0$, making (\ref{deft}) related to 
modular elliptic functions. It is worth mention that all $\vt$ functions 
fulfil the 1D diffusion equation, $\vt'' + c \dot \vt = 0$.

Let us recall the time $t$ dependent retarded Green's function (propagator):
\be\label{Green}
G({\bf r},t,{\bf r'},0) = \Theta(t)\sum_n \exp(-i E_n t) 
\Psi_n({\bf r}) \Psi_n^*({\bf r'})
\ee
where $\Theta(t)$ is Heaviside step function, $E_n$ are eigenvalues and 
$\Psi_n$ orthonormal complete eigenfunctions of the proper Hamiltonian and 
$\hbar = 1$. In our scaling, e. g. 3D HW box case, $E_n = \pi(i_x^2 + i_y^2 + 
i_z^2)$. One can go over to spectral representation performing a Fourier 
transformation
\begin{eqnarray}\label{Fou}
G({\bf r},{\bf r'},E) & = & -i\int_0^\infty \exp(iEt)G({\bf r},t,{\bf r'},0) dt
\nonumber \\ & = & \sum_n {\Psi_n({\bf r}) \Psi_n^*({\bf r'})\over E-E_n}
\end{eqnarray}
Inserting this into (\ref{bubble}) and exploiting the orthonormality one gets
the Meeron diagram value
\be
F_a(E) = C \sum_n {1\over (E-E_n)^2}
\ee
as it was done in \cite{LSF} and \cite{vanR} at $E=0$, see Eq. (\ref{varg}). 
The constant $C$, including appropriate combinatoric factor, will be 
specified later.

We have an alternative way. Let us integrate out the space variables
in advance
\be
Z(t) = \int d{\bf r} d{\bf r'} G({\bf r},0,{\bf r'},0) G({\bf r},t,{\bf r'},0) 
= \sum_n \exp(-i E_n t)
\ee
This simplifies the calculation, we will not have problems with the 
divergence of $\Pi({\bf r},{\bf r}) \propto S(y)$ in more than one dimension, 
or with non-uniformly convergent, ${\bf r}$-dependent series, Ref. \cite{Laco}.
Now we either first perform the Fourier transformation to get
$Z(E) = \sum (E-E_n)^{-1}$ and then apply a partial derivation, or we take 
the derivative of the Fourier integral to get
\be
F_a(0)= -C \frac {\partial Z(E)} {\partial E} \big\vert_{E=0} = C\int_0^\infty
y Z(y) dy
\ee
where $y$ now means imaginary time $y=it$, thus going over from (zero 
potential within a box) Schr\"odinger equation to the diffusion equation, 
or from Fourier to Laplace transformation. The quantity $Z(y)$ strongly 
resembles a partition function, with $y$ playing the role of inverse 
temperature. 

Imry \cite{Imry} also stated that variance of $g$ over random matrix 
ensembles can be calculated by a similar integral formula. Our version of 
(\ref{bubble}) and thus (\ref{varg}) with proper constant now reads:
\be\label{Ans0}
\langle g^2 \rangle _c = {3 F_a(0)\over 4} = {12 \over \pi^2} \int_0^{\infty}
y Z(y) dy.
\ee
Noticing that $Z(y)=\sum \exp(-E_n y)=-S'(y)/\pi$, after per parts 
we get an interesting, though for calculations not quite practical formula
\be
\langle g^2 \rangle _c = {12 \over \pi^3} \int_0^{\infty} S(y) dy.
\ee
Specifying $Z(y)$ in (\ref{Ans0}) with help of $\vt_3$ function 
(\ref{deft}) we can rewrite (\ref{varg}) for small integer dimensions 
$d$ as follows:
\be\label{Ans1}
\langle g^2 \rangle _c^{HW} = {12 \over 2^d \pi^2} \int_0^{\infty} y 
[\vt_3(0,q)-1][\vt_3(0,q)+1]^{d-1} dy.
\ee
This is easy to verify: each term with $\vt_3$ creates one summation
index in (\ref{HW}) for dimensions $d=1,2$ and 3. After per parts and 
integrating term by term we get (\ref{varg}). 

In 1D case, one can make use of the formula \cite{GR}
\be
\int_0^{\infty} x^{s-1} [\vt_3(0,e^{-\pi x^2})-1] dx = \pi^{-s/2} 
\Gamma(s/2) \zeta(s)
\ee
If we put $y=x^2$ in (\ref{Ans1}), specifying $s=4$ yields (\ref{U1D}).
\medskip

Similarly for PBC
\be\label{Ans2}
\langle g^2 \rangle _c^{PBC} = {6 \over \pi^2} \int_0^{\infty} y 
[\vt_3(0,q)-1][\vt_3(0,q^4)]^{d-1} dy.
\ee
Note that $q^4=\exp(-4\pi y)$.
Now we can calculate the UCF with high numerical precision for small integer
dimensions, see the Table 1. But the main generalization consists in regarding 
(\ref{Ans1}) and (\ref{Ans2}) as an analytical continuation in $d$, i. e. being
valid also for non-integer dimensions \cite{nas}  and even for those ones below 
1D, where it is hard to imagine any "perpendicular" direction. 

The Equations (\ref{Ans1}) and (\ref{Ans2}) represent the completed Ansatz. 
It reproduces the UCF for any integer dimension, i. e. any number 
of sums in (\ref{varg}), which should make the continuation unambiguous.
But this argument is rather naive, as the sums diverge for $d \ge 4$; some 
cut-off (e. g. as in \cite{Braun}) would be necessary. Note that the
integrals (\ref{Ans1}) and (\ref{Ans2}) also diverge for $d \ge 4$.
The dependence of UCF on dimension is plotted in Fig. 1. The common statement, 
that this dependence is weak \cite{vanR}, is true in the region $0\le d\le 3$.

We shall now  comment the limiting cases. For $d=0$ one gets sound values 
(see Table 1) both slightly below the universal constant 1/8, which corresponds 
to ballistic transport in quantum dots \cite{QDot}. For $d$ approaching 
4, $d=4-\varepsilon$ the leading contribution to integrals (\ref{Ans1}) and
(\ref{Ans2}) comes from small $y$ values and we can make use of the known 
alternative expansion:
\be\label{teta3}
\vt_3(u,\exp(-\pi x^2)) = {1\over x}\sum_{n=-\infty}^\infty
\exp\big[{(u-n\pi)^2\over \pi x^2}\big]
\ee
For small $x = \sqrt y$ and $u=0$ only the $n=0$ term is important 
(see Ref. \cite{Braun})
\be\label{t3}
\vt_3(0,\exp(-\pi x^2)) \approx {1\over x}
\ee
and we get
\be\label{expe}
\langle g^2 \rangle _c = {3\over 2\pi^2}{1\over\varepsilon} +{\cal O}
(1)
\ee
both for HW and PBC. These expansions remain valid even for anti-periodic BC,
if one simply replaces the second $\vt_3$ function in (\ref{Ans2}) and 
that one in (\ref{t3}) by $\vt_2$. The independence of the leading 
divergent term on boundary conditions seems to support our Ansatz. Similar 
statement was made in 3D for small $y$ expansion of $S_3(y)$, whose leading 
divergence also proved BC independent, see \cite{Braun}.

Unfortunately we cannot use formulas (\ref{Ans1}) and (\ref{Ans2}) to get 
the exact $L$-dependence of 4D conductance fluctuations directly. 
Introducing $1/L$ cut-off, e. g. by changing the lower limit of 
integration to small non-zero value, corresponds to making upper limits for 
summation indices in (\ref{varg}) finite. This would influence the propagator
used in one-loop diagrams so that the delta function on rhs of diffusion
equation in spectral representation (Refs. \cite{LSF}, \cite{vanR}) would 
become just a peak of finite, $L$ dependent width. Thus the summation 
of diagrams, mentioned in the beginning, would be no more simple.
Anyway, the too simple cut-off would give logarithmic $L$ dependence in 4D.
\medskip

In \cite{Peto} numerical calculations of UCF for 2D and 3D Anderson model 
were compared to theoretical values, computed also numerically 
with cca 3 digit precision from more complicated version of (\ref{varg}), 
given in \cite{LSF} and its PBC counterpart.
A plateau of $var\ g = \langle g^2 \rangle - \langle g \rangle ^2 = \langle g^2 
\rangle _c$ as a function of disorder $W$ was found with values reasonably 
close to theoretical predictions. The disorder had to be large 
enough to overcome the ballistic peak region, but not too close to the metal - 
insulator transition (if $ d > 2$). The plateau grew broader with increasing 
system size $L$.

We performed these calculations for 4D Anderson model with HW BC. Up to now 
only the region of metal - insulator transition was addressed in 4D, Refs. 
\cite{nas}, \cite{Gruss} and \cite{4D}. 
In diffusive regime, Ohm's law for cubic samples applies 
in the sense of leading contribution in $L$:
\be
\langle g \rangle = \sigma L^{d-2}
\ee
This serves as a test of diffusive kind transport, we expect $\langle g 
\rangle \propto L^2$, see Fig. 2. Here we find, that this
regime is well pronounced around $W=11$.

Results for $var\ g$ are shown in Fig. 3. A small plateau appears  
for larger $L$ and disorders cca within $10<W<11$. Typical statistical
ensembles are $N_{stat} = 10^5$ for $L \le 6$, $\approx 20000$ for $L=7$ and 
$\approx 3000$
for $L=8$. The value of the plateau diverges logarithmically (Fig. 4).

\bigskip

We conclude that a simple conjecture enables a high precision  calculation of 
UCF in any real dimension $0 \le d < 4$, with 
reasonable behavior in both limiting regions. Numerical calculations of 4D 
Anderson model show logarithmic divergence of CF with increasing 
system size.

\bigskip
\noindent Fruitful discussions with P. Marko\v s and Pavol Kalinay and carefull
reading of the manuscript are greatly acknowledged. This work was supported 
by Slovak Grant Agency VEGA 2/3108/23 and computer facilities of Slovak 
Academy of Sciences.

\begin{table}
\begin{center}
\begin{tabular}{llll}
\hline
\hline
$d$          & $\langle g^2 \rangle_{HW}$ & $\langle g^2 \rangle_{PBC}$ \\
\hline
\hline
 0    &   0.10439701  &    0.12291543  \\
 1    &   0.13333333  &    0.13333333  \\
 2    &   0.18561344  &    0.15407842  \\ 
 3    &   0.31405408  &    0.21939280  \\ 
\hline
 4    &   $\infty$    &    $\infty$  \\
\hline
\hline
\end{tabular}
\end{center}
\vspace*{2mm}
\caption{Numerical values of UCF, calculated by (\ref{Ans1}) and (\ref{Ans2}).
}
\end{table} 

\begin{figure}
\epsfig{file=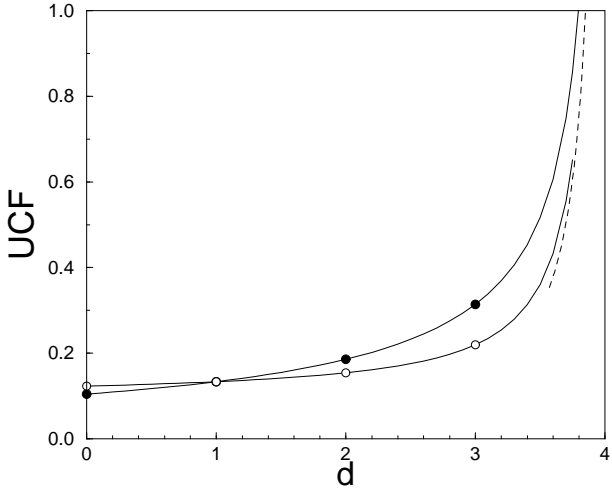,width=8cm}
\caption{ Universal conductance fluctuations as a function of dimension for
HW (open symbols) and PBC (full symbols). Dashed line is the asymptotic term 
from (\ref{expe}).
}
\end{figure}

\begin{figure}
\epsfig{file=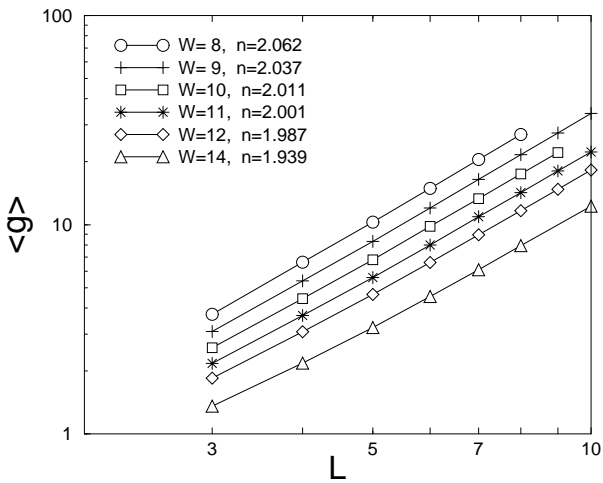,width=8cm}
\caption{ Mean conductance in 4D as a function of system size. Fitting 
parameter $n$ from $\langle g \rangle \propto L^n$  (for $L \ge 5$) shown 
in the Figure.
}
\end{figure}

\begin{figure}
\epsfig{file=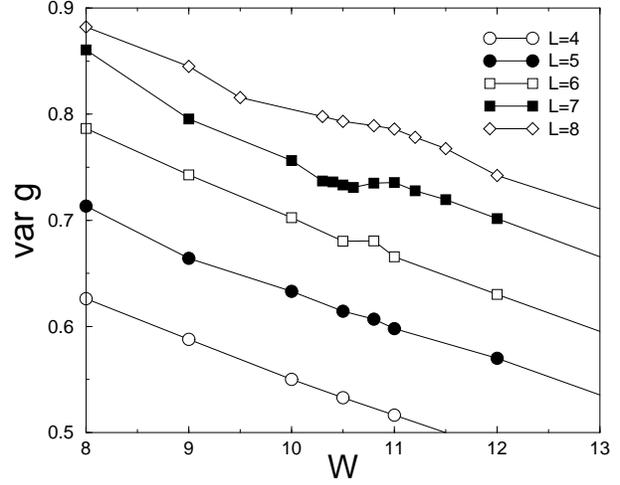,width=8cm}
\caption{ Conductance fluctuations in 4D as a function of disorder. System size
$L$ is described in the Figure.
}
\end{figure}

\begin{figure}
\epsfig{file=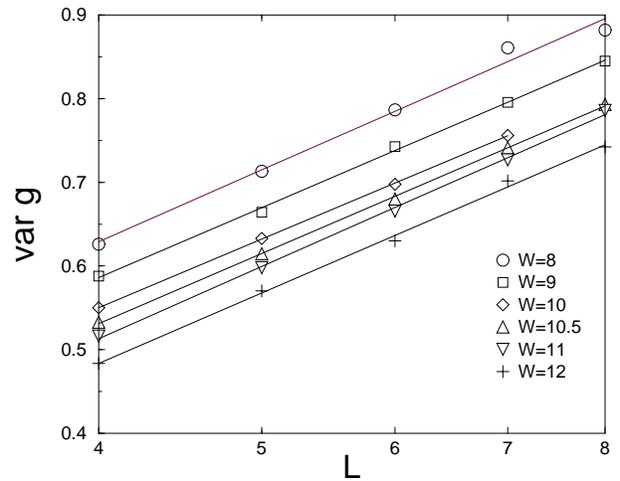,width=8cm}
\caption{ Conductance fluctuations in 4D as a function of $L$. The slope varies
from 0.36 to 0.39. 
}
\end{figure}

\end{document}